\documentstyle[twocolumn,aps]{revtex}
\input psbox
\begin{document}
\draft
\title{Metal-Insulator Transition in Two Dimensions in a Nearly Periodic
Potential} 
\author{K. Ziegler}
\address{Institut f\"ur Physik, Universit\"at Augsburg, 86135 Augsburg, 
Germany} 
\date{\today} 
\maketitle  
\begin{abstract}
A two-dimensional gas of non-interacting quasiparticles in a nearly periodic 
potential is considered at zero temperature. The potential is a superposition
of a periodic
potential, induced by the charge density wave of a Wigner crystal, and a weak
random potential due to disorder. There is a metal-insulator transition
that is controlled by the strength of the periodic potential. 
The transition is continuous in the presence of 
randomness.
We evaluate the density of states, which is non-zero at
the Fermi energy in the metallic phase, and the dc conductivity. The latter
changes with decreasing modulation of the periodic potential from 0 to
$\sigma\approx2e^2/h$.
\end{abstract} 
\noindent
The metal-insulator transition in silicon MOSFETs \cite{kravchenko} and
GaAs/AlAs heterostructures \cite{simmons} presents a new, unexpected
and interesting phenomenon in a two-dimensional electron (or hole) gas
\cite{kravchenko}.
One of the characteristic features of the transition is the scaling
behavior of the resistivity with temperature. There is a critical
resistivity which shows up at the transition point from the insulating
to the conducting regime. The corresponding critical conductivity typically
ranges from $0.3 e^2/h$ \cite{kravchenko} to $2e^2/h$ \cite{simmons}. 
In other words, it seems from the experiments that at zero temperature there
is either an insulating state or a conducting state with a conductivity
equal or larger than a critical (minimal) conductivity.
The nature of the conducting state (high electron density) is unclear.
It is either a normal metal \cite{castellani}, a superconducting state
\cite{belitz} or a state controlled by charged traps \cite{altshuler}.
The nature of the insulating state
(low electron density), on the other hand, is less controversal. Since the 2D
electron gas must have a high mobility in order to undergo a transition
to a non-insulating state, disorder is very weak in the samples.
Therefore, a Wigner crystal or at least a modulated electron density with
short range order is expected due to the Coulomb interaction \cite{yoon}.
In the following we shall discuss a model which
provides a simple picture of a metal-insulator transition in a nearly
periodic potential created by a modulated electron density.
The discussion will focus on the approach of the conducting
regime, coming from the insulating regime, but does not include a description
of this regime away from the transition point.

Lattice dynamics calculations \cite{meissner,bonsall} and computer
simulations \cite{ceperley,tanatar} for the pure 2D electron gas
indicate that the Wigner crystal forms a hexagonal lattice. 
Quasiparticles, which are the excitations in the
Wigner crystal, experience an effective potential due to the modulation
of the electron density. Formally,
the dynamics of the quasiparticles can be derived from the model of
a 2D electron gas which is subject to Coulomb interaction. Starting with
a microscopic model we can apply a self-consistent approximation for the
space-dependent electron density $n$ (cf. \cite{fradkin}), which describes
the Wigner crystal, and regard the fluctuations around the static electron
density as quasiparticles. In leading order the interaction of the
fluctuations is neglected, i.e. we consider independent quasiparticles.
This approach 
is analogous to the derivation of the Bogoliubov de Gennes Hamiltonian in
a superconductor, where the superconducting order parameter $\Delta$ is
treated in a self-consistent (BCS) approximation. In contrast to $\Delta$
the electron density $n$ couples directly to the quasiparticle density.
Therefore, it can be considered as an effective potential for the
quasiparticles. The latter prefer to stay in places with low electron
density (minimal potential, the circles shown in Fig. 1.)
but can also tunnel through the saddle points of the potential between
these potential minima (dashed lines in Fig. 1).
\begin{figure}
\begin{center}\mbox{\psbox{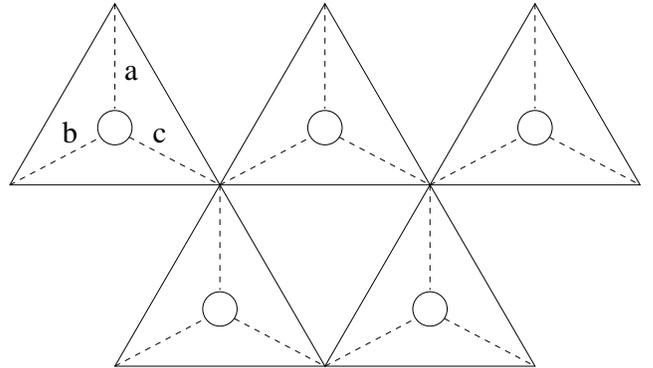}}\end{center}
\caption{Effective potential and the lattice of the tight-binding model.
Circles indicate minima of the electron density and dashed lines the
tunneling of the quasiparticles.}
\end{figure}

The model of the quasiparticles can be further simplified by assuming a
discrete potential instead of the smoothly varying density $n$. In other
words, the variation of $n$ is taken into account only down to a finite 
length scale. For the latter we choose the distance between a minimum and
an adjacent saddle point. This is a minimal model in order to study the
broken translational invariance caused by the density modulation.
It can be considered as a tight-binding approximation for the
quasiparticles, where the lattice points are the minima and the saddle
points of the electron density. There is a potential $V_{\bf R}$ at each
lattice site ${\bf R}$ with $V_{\bf R}=V_+$ on the minima and
$V_{\bf R}=V_-$ on the saddle points.
For the sake of simplicity of our model, (I) only the overlap between
nearest neighboring lattice sites (i.e. a minimum and a saddle point) 
and (II) no spin effect are taken into account.
It is convenient to substract a constant potential in order to
replace $V_{\bf R}$ by the modulation $m_{\bf R}\equiv V_{\bf R}
-(V_++V_-)/2$. Then $m_{\bf R}$ is a staggered field on the lattice with
$m_{\bf R}=\pm(V_+-V_-)/2\equiv\pm m$.
This field breaks the translational invariance on the lattice. A
translational-invariant representation can be obtained from one of
the two sublattices.
Using the sublattice of the maxima (circles in Fig. 1),
the co-ordinates of a site on this sublattice are ${\bf r}=(x,y)$, where
the $x$ ($y$) direction is horizontal (vertical) in Fig. 1.
The co-ordinates of a site of the other sublattice (corners of the
triangles in Fig. 1) are obtained by shifting the co-ordinates
${\bf r}$ with the unit vector ${\bf a}=(0,1)$. Any site ${\bf R}$ on the
lattice can now be written as $({\bf r},j)\equiv {\bf r}+(j-1){\bf a}$,
where $j=1,2$ is the index of the sublattice.

The quasiparticle Hamiltonian is defined as
\begin{equation}
{\hat H}=\sum_{{\bf r},{\bf r}'}\sum_{j,j'=1}^2
H_{{\bf r},j;{\bf r}',j'}c^\dagger_{{\bf r},j}c_{{\bf r}',j'},
\end{equation}
where $c$ ($c^\dagger$) are the annihilation (creation)
operators of the (fermionic) quasiparticles. 
For the off-diagonal matrix elements we assume $H_{{\bf r},{\bf r}+{\bf e}}
=t$,
where the unit vectors ${\bf e}={\bf a}, {\bf b}, {\bf c}$ are indicated in
Fig. 1. This Hamiltonian has also been studied in connection with the
quantum Hall effect. \cite{semenoff}
The matrix $H=H_0+m\sigma_3$, where the Pauli matrix $\sigma_3$ refers to
the sublattice index, can be given in Fourier representation with respect
to the sublattices (${\bf r}\to(k_x,k_y)$) as
\begin{equation}
H\to{\tilde H}=t[c(k_x,k_y)\sigma_1+s(k_x,k_y)\sigma_2]+m\sigma_3
\end{equation}
with Pauli matrices $\sigma_1$, $\sigma_2$ and
\begin{eqnarray}
c(k_x,k_y)=\cos(-k_y)+\cos(\sqrt{3}k_x/2+k_y/2)
\nonumber\\
+\cos(-\sqrt{3}k_x/2+k_y/2).
\end{eqnarray}
$s(k_x,k_y)$ is obtained from this expression by replacing the cosine by sine.
For this translational-invariant Hamiltonian we obtain the dispersion
\begin{equation}
E_{\pm}(k_x,k_y)=\pm\sqrt{m^2+t^2(c^2+s^2)}
\end{equation}
as shown in Fig. 2.
The modulation $m$ creates a gap between the two bands with $E_+>0$ and
with $E_-<0$. This gap vanishes only in the limit of a vanishing modulation
$m=0$. Thus our system is insulating as long as we have a modulated electron
density. In the case $m=0$ there are  six nodes $(k_x,k_y)$ on a circle
with radius $4\cdot3^{-3/2}\pi$, located at
\begin{equation}
(\pm {4\pi\over 3^{3/2}},0),\ \ ({2\pi\over 3^{3/2}},\pm{2\pi\over 3}),
\ \ (-{2\pi\over 3^{3/2}},\pm{2\pi\over 3}).
\end{equation}
For the analysis of the transport (i.e. low energy) properties of this system it
is sufficient to study the properties associated with the nodes. 
Expansion around the two nodes at $(\pm 4\pi/3^{3/2},0)$
leads to the Dirac Hamiltonians
\begin{equation}
H_0\sim \pm (3t/2)(k_1\sigma_1 \pm k_2\sigma_2),
\end{equation}
and around the other four nodes to
\begin{equation}
H_0\sim -(3t/4)[(\sqrt{3}k_1+k_2)\sigma_1+(\sqrt{3}k_1-k_2)\sigma_2],
\end{equation}
where $k_1$ ($k_2$) is now the deviation from the corresponding node in
$x$- ($y$-) direction. Thus for the pure system the Hamiltonian separates
into the six nodes when we consider small energy 
(i.e. large scale) behavior, where the vicinity of each node is given by a
two-dimensional Dirac operator.
\begin{figure}
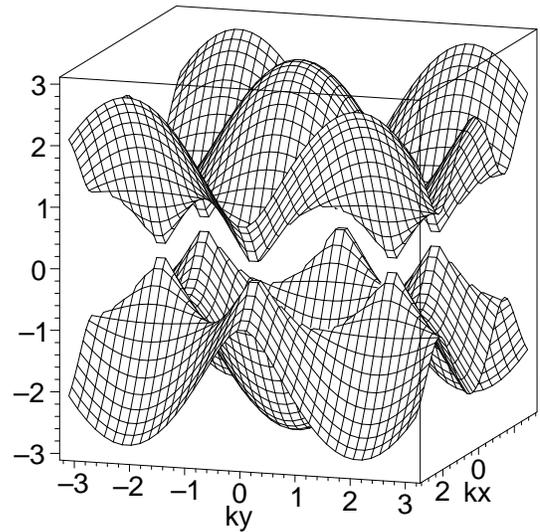

\begin{center}\mbox{\psbox{2dmitfig2.ps}}\end{center}
\caption{Dispersion $E(k_x,k_y)$ of quasiparticles, measured in units of
the overlap $t$.}
\end{figure}

So far we have assumed that there is perfect long-range order in the
Wigner crystal, leading to a translational-invariant Hamiltonian for
the quasiparticles. This is an idealization which is not valid in a real
system where we have always impurities. It is known that the 2D
Wigner crystal can be strongly affected by impurities.
In the presence of a high density of impurities the crystalline order
is destroyed completely and an electron glass can be created \cite{pastor}. 
On the other hand, short-range order may survive if disorder is only weak. 
This situation can be described in terms of the quasiparticles by
a weak random perturbation added to the periodic modulation. Various
types of disorder are possible 
but we will restrict the discussion to the simplest case, 
\begin{equation}
M_{\bf r}\sigma_3=(m+\delta M_{\bf r})\sigma_3,
\end{equation}
assuming that the effect of this type of randomness describes a generic
situation. 

In principle, the disorder $\delta M$ can couple different nodes because the
random term has fluctuations on all length scales. However, we will assume
that the overlap between the scales, related to the various nodes, is
negligible, i.e. for $k$ and $k'$ belonging to the vicinities of two 
different nodes we have
\begin{equation}
\sum_{\bf r}e^{-i{\bf k}\cdot {\bf r}+i{\bf k}'\cdot {\bf r}}
M_{\bf r}\approx 0.
\end{equation}
For each node we introduce an independent random term
$\delta M_{\alpha,{\bf r}}$
which means that $M_{\alpha,{\bf r}}$ fluctuates only on scales which are 
compatibel with the corresponding node. 
We assume a Gaussian distribution with $\langle\delta M_{\alpha,{\bf r}}
\rangle=0$ and $\langle\delta M_{\alpha,{\bf r}}
\delta M_{\alpha',{\bf r}'}\rangle =g
\delta_{\alpha,\alpha'}\delta_{{\bf r},{\bf r}'}$.
Then the Green's function $G(z)=(H-z)^{-1}$ can be
block-diagonalized with respect to the six nodes as
${\rm diag}[(H_1-z)^{-1},...,(H_6-z)^{-1}]$ with independent Dirac
Hamiltonians
\begin{equation}
H_\alpha=i(a_\alpha\nabla_1+b_\alpha\nabla_2)\sigma_1
+i(c_\alpha\nabla_1+d_\alpha\nabla_2)\sigma_2
+M_{\alpha,{\bf r}}\sigma_3,
\end{equation}
where $a_\alpha=-d_\alpha=-3t/2$, $b_\alpha=c_\alpha=0$ for the nodes with
$k_y=0$ (i.e. $\alpha=1,2$)
and $a_\alpha=c_\alpha=-3^{3/2}t/4$, $b_\alpha=-d_\alpha=-3t/4$ for the
nodes with $k_y\ne0$ (i.e. $\alpha=3,...,6$). 
The block-diagonal structure is a crucial simplification because we
only have to evaluate the physical quantities for the six nodes
independently. The density of states (DOS) and the conductivity of the 
corresponding Dirac Hamiltonians are known and shall be discussed
subsequently.

\noindent
The DOS of a system near a metal-insulator transition exhibits a
characteristic behavior. In the simplest case there is a gap in the
insulating phase
which closes at the transition to the metal. The closing of the gap
was measured recently in a tunneling experiment near the metal-insulator
transition of boron-doped silicon crystals, a three-dimensional electron gas.
\cite{lee} 
It was found that the DOS goes roughly like a power law
$\rho(E)\sim\rho_0 (|E|/E_0)^\gamma$ ($E$ is the distance from the
Fermi energy $E_F$ and $\gamma\approx 0.5$) in the insulating phase and like
$\rho(E)\sim\rho_0 [1+(|E|/E_0)^\gamma]$ in the metallic
phase. Thus the insulating phase can be distinguished from the metallic
phase only for energies close to the Fermi energy. A similar behavior is
expected for a two-dimensional system, with a different exponent $\gamma$
though. For instance, a linear behavior $\rho(E)\sim\rho_0 |E|/E_0$
was observed in a 2D electron gas subject to a perpendicular magnetic field.
\cite{chan}
This result is in agreement with the effective Dirac fermion description of
quasiparticles in a quantum Hall system \cite{semenoff} which also
gives a linear DOS in the absence of disorder.
Moreover, a linear behavior of the DOS was found in a mean-field
calculation of a disordered 2D electron gas. \cite{pastor}

In our model, the block-diagonal structure enables us to
write the DOS as a sum of DOS's of the
vicinity of each node. Using the result of the Dirac
fermions from a saddle point approximation this gives for $m=0$ at
each node $\alpha$ \cite{eckern}
\begin{equation}
\rho_\alpha(E)\approx {e^{-\pi/g}\over g}+|E|^\gamma,
\end{equation}
where the exponent $\gamma$ varies with increasing randomness $g$
from 1 to 0 (Fig. 3).
\begin{figure}
\begin{center}\mbox{\psbox{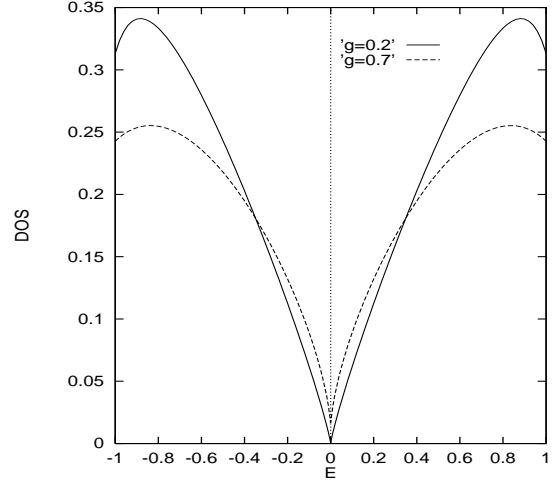}}\end{center}
\caption{Density of states at $m=0$ for different strength of randomness}
\end{figure}
The energy scale is set by the overlap $t$.
The DOS as a function of the modulation parameter $m$
at $E=0$ is $\propto\sqrt{4e^{-2\pi/g}-m^2}\Theta(4e^{-2\pi/g}-m^2)/2g$.
Therefore, the DOS vanishes with $g\to0$. 
The effect of the periodic modulation is the creation of a gap.
On the other hand, the random fluctuations around the periodic 
modulation create states at the Fermi energy as long as the
random fluctuations are strong relative to the periodic modulation.
To have quasiparticle states at the Fermi energy it requires random 
fluctuations, characterized by
the parameter $2e^{-\pi/g}$, which are larger than the periodic
modulation, characterized by $m$, . The creation of new states
around $m=0$ by the randomness can be explained by the formation of
tail states at low energy. This is similar to the creation of Lifshitz
tails. \cite{lifshits}
However, it will be discussed in the following that these states
are conducting - in contrast to the localized states in the Lifshitz tails.


In the presence of a gap the conductivity vanishes with vanishing
temperature.
Here we are only interested in zero temperature properties, i.e. we
must study the transport properties for the system with a closed gap.
Since a randomly disturbed modulation ($\delta M_r\ne0$) creates
states at the Fermi level for $m\le 2e^{-\pi/g}$, the case of the nearly
periodic modulation is a candidate for a metal-insulator transition.
However, the vanishing gap is necessary but not sufficient for a conducting
state because the quasiparticle states might be localized by randomness.
The localization effect is known to be very efficient in $d=2$, e.g.,
all quantum states are localized according to conventional scaling theory
for localization \cite{abrahams}. However, it was found that Dirac
fermions with random mass can escape from localization by a
special mechanism due to symmetry breaking \cite{ziegler972}. 
It leads
to diffusion of the Dirac particle for sufficiently small average mass,
i.e. for sufficiently small modulation.
The conductivity can be evaluated using the Kubo-Greenwood formula.
Then the dc conductivity at $T=0$ and $E=0$ reads
\begin{eqnarray}
\sigma\approx{e^2\over h}
\lim_{\eta\to0}\eta^2\sum_{\bf r} {\bf r}^2{\rm Tr}_2
\langle G(0,{\bf r};i\eta)G({\bf r},0;-i\eta)\rangle
\nonumber\\
=-{e^2\over h}\lim_{\eta\to0}\eta^2\nabla_k^2
{\tilde C}(k,\eta)|_{k=0}
\label{cond}
\end{eqnarray}
with the two-particle Green's function $C({\bf r},\eta)={\rm Tr}_2
\langle G(0,{\bf r};i\eta)G({\bf r},0;-i\eta)\rangle$.
The latter is a sum of contributions from the six different nodes due to the
block-diagonal form of the single-particle Green's function. The average
two-particle Green's function can be diagonalized by a Fourier transformation
because it is translational invariant as a consequence of the uniform
distribution of the randomness.
The nodes require a similarity transformation
\begin{eqnarray}
k_1\to k_1'=a_\alpha k_1+b_\alpha k_2, \ \ k_2\to 
k_2'=c_\alpha k_1+d_\alpha k_2
\end{eqnarray}
in order to obtain the usual Dirac form $\sigma\cdot k$ for the
Hamiltonian. 
The average two-particle Green's function is \cite{ziegler972}
\begin{eqnarray}
{\tilde C}_\alpha(k,\eta)=
{\pi\over2}{\rho\over \eta+D({k'_1}^2+{k'_2}^2)
/|a_\alpha d_\alpha-b_\alpha c_\alpha|},
\end{eqnarray}
where $D$ is the diffusion coefficient of the Dirac fermions. The extra factor
$1/|a_\alpha b_\alpha -c_\alpha d_\alpha|$ is the Jacobian of the
transformation $k\to k'$.
For weak disorder ($g\ll 1$) the diffusion coefficient is \cite{jug}
\begin{equation}
D\approx{1\over2\pi^2\rho}(1-x^2)[1-{g\over2\pi}(1-2x^2)]\Theta(1-x^2)
\label{diff}
\end{equation}
with $x=me^{\pi/g}/2$. With (\ref{cond}) and the summation over all nodes
this implies for the conductivity 
\begin{eqnarray}
\sigma(m)\approx {e^2\over h}s_0(1-x^2)[1-{g\over2\pi}(1-2x^2)]\Theta(1-x^2)
\end{eqnarray}
with $s_0\approx 2.1$. The conductivity depends on the modulation
$m$ which scales with $e^{\pi/g}$. The transition is very sharp for weak
disorder, since the scale is exponential, and it is discontinuous with
a minimal conductivity $2e^2/h$ if disorder is absent.

The quasiparticle states overlap in such a manner
that they form conducting states. This effect is known for Dirac
fermions with randomness already in one dimension, where the zero
energy modes are extended. \cite{lifshits} However, in contrast
to the finite diffusion coefficient (\ref{diff}), leading to a
diffusive behavior of quasiparticles, this coefficient is singular then.
In our model
the diffusion is valid only for weak randomness, as it is indicated
by the vanishing diffusion coefficient (\ref{diff}) if $g=2\pi$.

In conclusion, a possible metallic state in a two-dimensional
electron gas is approached from an insulating regime by the destruction of
the gap of the insulating state by disorder.
Our discussion is based on a simple model for quasiparticles in a nearly
periodically modulated electron density, which can be considered as a Wigner
solid with weak disorder. The system becomes metallic for a sufficiently
weak modulation with a conductivity $\sigma\approx2e^2/h$. The latter
decreases linearly with increasing disorder. The conductivity varies
continuously for the nearly periodic potential.

\noindent
Acknowledgement: This work was supported by the Sonderforschungsbereich
484 of the Deutsche Forschungsgemeinschaft.

\end{document}